\documentclass[showkeys,superscriptaddress,nofootinbib,floatfix,prd,10pt,aps]{revtex4-2}
\usepackage{graphicx,epstopdf}
\usepackage[caption=false]{subfig}
\usepackage{appendix}
\usepackage[T1]{fontenc}
\usepackage{lmodern}
\usepackage[dvipsnames,x11names]{xcolor}
\usepackage[colorlinks=true,linkcolor=NavyBlue,citecolor=ForestGreen,urlcolor=NavyBlue]{hyperref}
\usepackage[sort&compress]{natbib}
\usepackage{lipsum}
\usepackage{morefloats}
\usepackage[pdf]{pstricks}
\usepackage{amsmath}
\usepackage{amssymb}
\usepackage{amsfonts}
\usepackage{rotating}
\usepackage{cancel}
\usepackage{mathtools}
\usepackage{bbm}
\usepackage{dsfont}
\usepackage{bbold}
\usepackage{multirow}
\usepackage{ulem}
\usepackage{physics}
\usepackage{orcidlink}
\usepackage{colortbl}
\usepackage{enumitem}
\usepackage{subfloat}
\usepackage{booktabs}

\begin{document}

\title{Magnetic Moments of Hidden-Charm Pentaquarks in the Diquark-Diquark-Antiquark Scheme}

\author{Halil Mutuk\orcidlink{0000-0002-6794-0879}}%
\email[]{hmutuk@omu.edu.tr }
\affiliation{ Department of Physics, Faculty of Sciences, Ondokuz Mayis University, 55200, Samsun, T\"{u}rkiye}

\begin{abstract}
The magnetic moment of a hadron is an important spectroscopic parameter that encodes valuable information about its internal structure. In this work, we systematically investigate the magnetic moments of hidden-charm pentaquark states, including the experimentally observed $P_c(4457)$ and related configurations with and without strangeness. The analysis is performed within the diquark-diquark-antiquark framework for spin-parity quantum numbers $J^P = \frac{1}{2}^-$, $\frac{3}{2}^-$, and $\frac{5}{2}^-$. Magnetic moment values are computed for different spin and flavor configurations, and the results are compared with existing predictions in the literature. These predictions may offer insight into the inner structure and quantum numbers of these exotic states, and potentially help distinguish between different theoretical models.
\end{abstract}

\maketitle

\section{Introduction}\label{motivation}

Quark model which was introduced by Gell-Mann and Zweig independently \cite{Gell-Mann:1964ewy,Zweig:1964ruk} describes mesons as bound states of a quark and an antiquark, and baryons as bound states of three quarks. Quark model have successfully explained physical features of experimentally observed meson and baryon states till to the first years of 2000s. The paradigm has changed when the observation of $\chi_{c1}(3872)$ (also known as $X(3872)$) was announced, a state that has unusual properties which cannot be fitted to the quark model \cite{Belle:2003nnu}. Since then many states have been reported which cannot be collected as ordinary mesons $(q \bar q)$ or baryons $(qqq)$. These are called exotic states, neither quark model nor quantum chromodynamics (QCD) prohibit the existence of exotic states. 

QCD is the quantum field theory of strong interaction, not only governs the interactions between quarks and gluons but also the interaction between color-neutral hadrons. Short and long distance behaviours of QCD is of primary interest since perturbative and nonperturbative interactions are related to these distances. Investigating hadron-hadron interactions based on QCD presents opportunities for multiquark states. For example, four-quark states can be studied either compact tetraquark or molecular states. In both cases, physical interpretations can lead to different physical results. For this reason, unravelling internal structure of the multiquark states is one of the major topic in hadron physics. 

To date back, the existence of multiquark states were conjectured in the original works of Gell-Mann and Zweig. They showed that it is possible to obtain color-neutral objects by combining four, five or even six quarks. Among these multiquark states, five quark states (namely pentaquarks) are important members of multiquark states. In 2015, LHCb Collaboration reported the observation of pentaquark states. $P_c(4380)$ and $P_c(4450)$ states were confirmed in the $J/\psi+p$ decay channel. Four years later, LCHb Collaboration accumulated tied more data on these states which yielded important results. Previously reported $P_c(4450)$ state had split into $P_c(4440)$ and $P_c(4457)$ states, and another resonance, named as $P_c(4312)^+$ had been discovered. The status of the other previously reported $P_c(4380)$  state remains unresolved: neither confirmed nor refuted. Following the year 2019, in 2020 LHCb Collaboration observed a pentaquark state, $P_{cs}(4459)$, in the invariant mass spectrum of $J/\psi\Lambda$ in the $\Xi_b^0 \rightarrow J/\psi\,\Lambda\,K^-$ decay. In 2022, the LHCb collaboration announced observation of a new structure $P_{cs}(4338)$ in the $J/\psi\Lambda$ mass distribution of the $B^- \rightarrow J/\psi\Lambda^- p $ decays. The spectroscopic parameters, minimal valence quark contents, and observed channels for these states are presented in Table \ref{pentaquarks}.

\begin{widetext}

\begin{table}[htb]
\caption{Hidden-charm pentaquark states observed by the LHCb Collaboration.}\label{pentaquarks}
\begin{tabular}{l|c|c|c|c}
\toprule
State  & Mass (MeV) & Width (MeV) & Content &  Observed channels\\
\toprule
$P_c(4380)^+$ \cite{LHCb:2015yax}  &            $4380\pm8\pm29$          &       $215\pm18\pm86$           & $uudc\bar{c}$ & $\Lambda_b^0 \to J/\psi pK^-$\\
$P_c(4312)^+$ \cite{LHCb:2019kea}       & ~~$4311.9\pm0.7^{~+6.8}_{~-0.6}$~~  &  $9.8\pm2.7^{~+3.7}_{~-4.5}$    & $uudc\bar{c}$ & $\Lambda_b^0 \to J/\psi pK^-$ \\
$P_c(4440)^+$ \cite{LHCb:2019kea}      &    $4440.3\pm1.3^{~+4.1}_{~-4.7}$   &  $20.6\pm4.9^{~+8.7}_{~-10.1}$  & $uudc\bar{c}$& $\Lambda_b^0 \to J/\psi pK^-$ \\
$P_c(4457)^+$  \cite{LHCb:2019kea}    &    $4457.3\pm0.6^{~+4.1}_{~-1.7}$   &  $6.4\pm2.0^{~+5.7}_{~-1.9}$    & $uudc\bar{c}$ & $\Lambda_b^0 \to J/\psi pK^-$ \\
$P_{cs}(4459)^0$ \cite{LHCb:2020jpq}   &    $4458.8\pm2.9^{~+4.7}_{~-1.1}$   &  $17.3\pm6.5^{~+8.0}_{~-5.7}$   & $udsc\bar{c}$ &~$\Xi_b^- \to J/\psi \Lambda K^-$ \\
$P_{cs}(4338)^0$ \cite{LHCb:2022ogu}  &    $4338.2 \pm 0.7 \pm 0.4$   &  $7.0 \pm 1.2 \pm 1.3$    & $udsc\bar{c}$ & $B^-\to J/\psi \Lambda \bar{p}$\\
\toprule
\end{tabular}
\end{table}

\end{widetext}

The observation of the aforementioned pentaquark states triggered many theoretical studies with various methods to elucidate inner structure, exact nature, and determine quantum numbers of these nonconventional states. The pentaquark states consist of five quarks and clustering these quarks into substructures seems reasonable. The popular schemes are (i) molecular scheme, (ii) diquark-diquark-antiquark scheme, and (iii) diquark-triquark scheme. These pictures have different quark rearrangements. The primary assumptions of these pictures are as follows:
\begin{enumerate}[label=(\roman*)]
\item Molecular state scheme: In this scheme, pentaquark state consists of a meson and a baryon. Color neutral constituents (meson and baryon) form color singlet pentaquark state in molecular configuration.

\item Diquark-diquark-antiquark scheme: In this case, diquark prefers to form color antitriplet $\bar{3}_c$. Two color antriplet diquarks together with color antitriplet quark $\bar{3}_c(\text{diquark)} \otimes \bar{3}_c(\text{diquark)} \otimes \bar{3}_c(\text{antiquark)} $ form color singlet pentaquark. 

\item Diquark-triquark scheme: This scheme has a similar configuration with respect to molecular scheme. In this type of configuration, triquark consists two quarks and antiquark. Color triplet triquark $3_c(\text{triquark)}$ and color antitriplet diquark $\bar{3}_c(\text{diquark)}$ form diquark-triquark configuration $3_c(\text{triquark)} \otimes \bar{3}_c(\text{diquark)}$ as color singlet a pentaquark state. 

\end{enumerate}


The above models are frequently used to study spectroscopic parameters of the hidden-charm pentaquark states. Although the literature consist many studies about hidden-charm pentaquark states, mostly about mass spectra, their internal nature and quantum numbers are still incomplete. However spectroscopic parameters alone may be inadequate to enlighten the controversial nature of these states. Consequently, to unveil the internal structure and configurations of these states, further studies including magnetic moments, radiative decays and weak decays are required. Among these, the information that magnetic moment prevails is essential to gain insight into the nature and internal structure of exotic states, which almost all of them are controversial in terms of internal structure. The magnetic moment is a measure of the distribution of quark-antiquark pairs within the hadron. It is also related to the shape of the hadron, electric radii and magnetic radii of the hadrons. In addition to this, pentaquark magnetic moments have a significant effect on differential and total cross-sections in electroproduction and photoproduction of pentaquarks. In this respect, magnetic moment is a crucial ingredient in calculation of cross sections of $J/\psi$ photoproduction which can provide observation of further pentaquark states \cite{Bijker:2004gr,Ortiz-Pacheco:2018ccl}. 

\begin{figure}[h!]
\centering
\includegraphics[width=8.5cm]{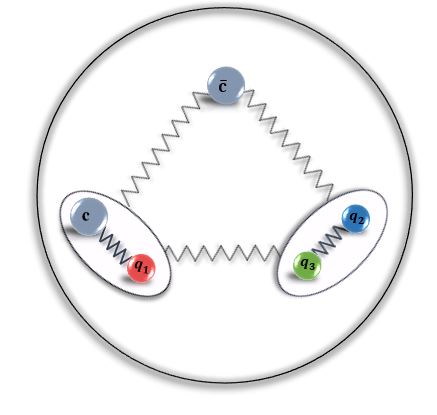}
\caption{Schematic diagram of pentaquark in diquark-diquark-antiquark scheme.}
\label{ddaconfig}
\end{figure}

In this present work, we adopt diquark-diquark-antiquark picture for the  $P_{c}(4457)$ and related hidden-charm pentaquark states with and without strangeness and study magnetic moments. This picture can be visualized in Figure \ref{ddaconfig}.

The diquark-diquark-antiquark model provides a framework for understanding the structure and properties of these pentaquark states. This model has been instrumental in describing the masses and decay properties of various observed states. For instance, the LHCb data on pentaquarks with open charm emphasize the importance of hadron components in their structure, suggesting that these states can be described both in terms of quarks and hadrons \cite{Anisovich:2017aqa}. It is pointed out in Ref. \cite{Barabanov:2020jvn} that the development of lattice-regularized QCD and new theoretical approaches to the continuum bound-state problem have strongly suggested the importance of soft quark+quark (diquark) interactions in hadron physics. A recent LHCb analysis on the nature of $\chi_{c1}(3872)$ reveals that the radiative decays of $\chi_{c1}(3872)\to \psi(2S) \gamma $ and $\chi_{c1} (3872)\to J/\psi \gamma $ indicate that $D^0 \bar{D}^{\ast 0} + \bar{D}^0 D^{\ast 0}$ picture for the  $\chi_{c1}(3872)$ is questionable and favors a sizeable compact component in the $\chi_{c1}(3872)$ state \cite{LHCb:2024tpv}. The other point is that diquark-triquark model has a similar quark configuration as the molecular model has, the exception is just the color representation. 

Despite the fact that magnetic moment gives valuable information on physical parameters that are directly correlated with the inner structure of the state, the number of studies related to magnetic moment are not at the desired level \cite{Wang:2016dzu,Ozdem:2018qeh,Ortiz-Pacheco:2018ccl,Xu:2020flp,Ozdem:2021ugy,Li:2021ryu,Gao:2021hmv,Ozdem:2022iqk,Wang:2022nqs,Wang:2022tib,Wang:2023iox,Guo:2023fih,Wang:2023ael,Ozdem:2024jty,Li:2024wxr,Li:2024jlq,Ozdem:2024rqx,Mutuk:2024ltc,Mutuk:2024jxf,Lai:2024jfe}. Although the magnetic moments of the hadrons give valuable information about the internal structure, magnetic moment measurement is a real challenge. Magnetic moment is the response of a state to an external magnetic field and if the particle lifetime is relatively short, it is hard to measure. However, magnetic moment measurement is possible indirectly, provided sufficient data is accumulated. For example, $\Delta^+(1232)$ baryon has also a very short lifetime, $(\sim 5 \times 10^{-24} \ \text{s})$, however its magnetic moment was obtained through $\gamma N $ $ \rightarrow $ $ \Delta $ $\rightarrow $ $ \Delta \gamma $ $ \rightarrow$ $ \pi N \gamma $ process  \cite{Pascalutsa:2004je, Pascalutsa:2005vq, Pascalutsa:2007wb}. A comparable process may be employed to achieve the magnetic moment of the $P_{c}$ pentaquarks. QCD based models can also be employed to obtain magnetic moments. In Refs. \cite{Can:2013zpa,Can:2013tna}, magnetic moments of baryons containing two charm quarks have been extracted by lattice QCD formalism. These analyses can be extended to the exotic hadrons.

We organize this paper in the following manner: In Sec.~\ref{formalism} we present the method which is used to calculate the magnetic moment of the considered states, named as $P_c$.  Sec.~\ref{numerical} is dedicated to numerical results of the magnetic moments of the $P_c$ states. Finally, Sec.~\ref{summary} is reserved for summary of this work.

\section{Theoretical formalism}\label{formalism}

\subsection{Wave function}
 The total wave function of a hadronic state depend on both spatial degrees of freedom and the internal degrees of freedom of color, flavor and spin
\begin{equation}
\Psi_{\text{total}}=\phi_{flavor}\chi_{spin}\epsilon_{color}\eta_{spatial}.
\end{equation}
The wave function of a pentaquark state can be constructed by imposing symmetry considerations: (i) the total wave function of pentaquark should be antisymmetric, and (ii) all physical states should be colorless, i.e., color neutral. 

According to the color configurations, the pentaquark state may consist of either two or three clusters. In our model, pentaquark is composed of three clusters: two diquarks plus an antiquark. In color space, a diquark is in the $\bar{3}_c$ representation of the $SU(3)$ color group. Introducing the notation 
\begin{equation}
(S_t, R_f, R_c),
\end{equation} 
where, $S_t$ denotes the spin of two quarks, $R_f$ and $R_c$ are the irreducible representations of two-quark states in flavor $\text{SU}_f(3)$ and color $\text{SU}_c(3)$ spaces, respectively; two quarks in S-wave can have the following configurations
\begin{equation}
(0, 6_f, 6_c), ~(1, \bar{3}_f, 6_c), ~(0, \bar{3}_f, \bar{3}_c), ~(1, 6_f, \bar{3}_c). \label{eq3}
\end{equation}

As noted in Eq. (\ref{eq3}), diquarks can form symmetric $(6_c)$ or antisymmetric $(\bar{3}_c)$ color configurations. These configurations are allowed by color SU(3). The antisymmetric $(\bar{3}_c)$ is chosen because it is the dominant configuration in most phenomenological models of compact pentaquarks, as it allows color-singlet formation with the antiquark. It is also energetically favored due to the attractive nature of the color interaction in this channel which is a well-established result in QCD-based models and supported by lattice QCD simulations and phenomenological studies (see Refs. \cite{Anisovich:2017aqa,Barabanov:2020jvn}). The symmetric 
$(6_c)$ diquark would require additional color-neutral binding mechanisms (e.g., gluon exchanges), complicating the model without clear experimental justification.

In spin-flavor space, two quarks inside a diquark should be symmetric in order to satisfy Fermi statistics, due to the fact that they are antisymmetric in color space. The possible diquark configurations are then $(0, \bar{3}_f, \bar{3}_c)$ and $(1, 6_f, \bar{3}_c)$. As is clear from these representations, a diquark prefers to form $\bar{3}_c$ in color space and $6_f$ or $\bar{3}_f$ in flavor space. The flavor wave function of the hidden-charm pentaquark states can be constructed as follows in the notation of $\phi(\text{Flavor representation}, \text{Isospin}, \text{Strangeness})$
\begin{align}
&\phi(8_{1f},1/2,0) = -\left\{qq\right\} (qc) \bar{c}, \label{flavorwavebegin} \\ 
&\phi(8_{1f},0,-1) = -\left\{qs\right\} (qc)\bar{c},   \\
&\phi(8_{1f},1,-1) = \sqrt{\frac{2}{3}} \left\{qq\right\} (sc) \bar{c} -\sqrt\frac{1}{3} \left\{qs\right\} (qc) \bar{c},  \\
&\phi(8_{1f},1/2,-2) = \sqrt\frac{1}{3} \{qs\} (sc) \bar{c} - \sqrt{\frac{2}{3}} \{ss\}(qc) \bar{c}, \\
& \phi(8_{2f},1/2,0) = [qq] (qc) \bar{c},   \\
& \phi(8_{2f},1/2,-2) = [qs] (sc) \bar{c},  \\
& \phi(8_{2f},0,-1) = -\sqrt{\frac{2}{3}} [qq] (sc) \bar{c} - \sqrt\frac{1}{3} [qs] (qc) \bar{c},  \\
& \phi(8_{2f},1,-1) =  [qs] (qc) \bar{c}, \\
& \phi(10_f,3/2,0) = \{qq\} (qc) \bar{c}, \\
& \phi(10_f,1,-1) = \sqrt\frac{2}{3} \{qs\} (qc)\bar{c} + \sqrt\frac{1}{3} \{qq\}(sc)\bar{c}, \\
& \phi(10_f,1/2,-2) = \sqrt\frac{2}{3} \{qs\} (sc) \bar{c} + \sqrt\frac{1}{3} \{ss\} (qc) \bar{c},  \\
& \phi(10_f,0,-3) = \{ss\} (sc) \bar{c}. \label{flavorwaveend}
\end{align}
Here $q$ represents the light quarks, $u$ and  $d$. In the above flavor wave functions, the bracket $``\left\{~\right\}"$ for the first diquark denotes that two quarks are symmetric (spin 1) inside it and the bracket $``\left[~\right]"$ for the first diquark denotes two quarks are antisymmetric (spin 0) inside it in the SU(3) flavor space. For the second diquark, the bracket $``(~)"$ means that flavor symmetry of the two quarks inside this diquark is not definite. It is possible to use SU(2) Clebcsh-Gordon coefficients to expand the flavor wave functions in Eqs. (\ref{flavorwavebegin}-\ref{flavorwaveend}) to obtain flavor wave functions in $8_f$ representation (Table \ref{pentaquarks8f}) and $10_f$ representation (Table \ref{pentaquarks10f}):

\begin{widetext}

\begin{table*}[htb]
\caption{Flavor wave functions of $P_c$ and related states in $8_f$ representation}\label{pentaquarks8f}
\begin{tabular}{l|c|c}
\toprule
$(I,I_3)$ & Wave function-$8_{1f}$& Wave function-$8_{2f}$  \\
\toprule
$(\frac{1}{2},\frac{1}{2})$ &    $-\sqrt{\frac{1}{3}}\{ud\}(cu){\bar c}+\sqrt{\frac{2}{3}}\{uu\}{({c}d)\bar c}$       &   $[ud]({c}u)\bar c$          \\
$(\frac{1}{2},-\frac{1}{2})$   & $\sqrt{\frac{1}{3}}\{ud\}(cd){\bar c}-\sqrt{\frac{2}{3}}\{dd\}({c}u){\bar c}$ &   $[ud](cd){\bar c}$   \\
$(1,1)$   &  $\sqrt{\frac{1}{3}}\{us\}({c}u){\bar
c}-\sqrt{\frac{2}{3}}\{uu\}(cs){\bar c}$  &  $[us](cu){\bar c}$    \\
$(1,0)$  &    $\sqrt{\frac{1}{6}}[\{us\}({c}d){\bar
c}+\{ds\}({c}u){\bar c}]-\sqrt{\frac{2}{3}}\{ud\}({c}s){\bar c}$  &  $\frac{1}{\sqrt2}\{ [us](cd){\bar c}+[ds](cu){\bar c} \}$    \\
$(1,-1)$& $ \sqrt{\frac{1}{3}}\{ds\}({c}d){\bar
c}-\sqrt{\frac{2}{3}}\{dd\}(cs){\bar c}$
 &  $[ds](cd){\bar c}$     \\
$(0,0)$  & $\sqrt{\frac{1}{2}}[\{ds\}(cu){\bar
c}-\{us\}(cd){\bar c}]$   &  $\frac{1}{\sqrt6} \{ [us](cd){\bar c}-[ds](cu){\bar c}-2[ud](cs){\bar c} \}$ \\
$(\frac{1}{2},\frac{1}{2})$ & $\sqrt{\frac{1}{3}}\{us\}({c}s){\bar
c}-\sqrt{\frac{2}{3}}\{ss\}({c}u){\bar c}$   &  $[us](cs){\bar c}$  \\

$(\frac{1}{2},-\frac{1}{2})$ &$\sqrt{\frac{1}{3}}\{ds\}({c}s){\bar
c}-\sqrt{\frac{2}{3}}\{ss\}({c}d){\bar c}$   &  $[ds](cs){\bar c}$ \\

\toprule
\end{tabular}
\end{table*}

\end{widetext}

\begin{widetext}

\begin{table*}[htb]
\caption{Flavor wave functions of $P_c$ and related states in $10_f$ representation}\label{pentaquarks10f}
\begin{tabular}{l|c}
\toprule
$(I,I_3)$ & Wave function-$10_{f}$  \\
\toprule
$(\frac{3}{2},\frac{3}{2})$ &    $\{uu\}(cu){\bar c}$             \\
$(\frac{3}{2},\frac{1}{2})$   & $\sqrt{\frac{2}{3}}\{ud\}(cu){\bar c}+\sqrt{\frac{1}{3}}\{uu\}(cd){\bar c}$  \\
$(\frac{3}{2},-\frac{1}{2})$  &   $\sqrt{\frac{2}{3}}\{ud\}(cd){\bar c}+\sqrt{\frac{1}{3}}\{dd\}(cu){\bar c}$    \\
$(\frac{3}{2},-\frac{3}{2})$  &   $\{dd\}(cd){\bar c}$    \\
$(1,1)$&  $\sqrt{\frac{2}{3}}\{us\}(cu){\bar c}+\sqrt{\frac{1}{3}}\{uu\}(cs){\bar c}$  \\
$(1,0)$  & $\sqrt{\frac{1}{3}}[\{us\}(cd){\bar c}+\{ds\}(cu){\bar c}]+\sqrt{\frac{1}{3}}\{ud\}(cs){\bar c}$   \\
$(1,-1)$ & $\sqrt{\frac{2}{3}}\{ds\}(cd){\bar c}+\sqrt{\frac{1}{3}}\{dd\}(cs){\bar c}$  \\
$(\frac{1}{2},\frac{1}{2})$ & $\sqrt{\frac{1}{3}}\{ss\}(cu){\bar c}+\sqrt{\frac{2}{3}}\{us\}(cs){\bar c}$   \\
$(\frac{1}{2},-\frac{1}{2})$ & $\sqrt{\frac{1}{3}}\{ss\}(cd){\bar c}+\sqrt{\frac{2}{3}}\{ds\}(cs){\bar c}$   \\
$(0,0)$ & $\{ss\}(cs){\bar c}$  \\
\toprule
\end{tabular}
\end{table*}

\end{widetext}

In spin space, the wave function of the pentaquark can be written as
\begin{equation}
\vert S_{12},S_{34},S_{1234},S \rangle = \vert \left\{\left[\left(s_1s_2\right)_{S_{12}} \otimes \left(s_3s_4\right)_{S_{34}} \right]_{S_{1234}} \otimes s_{\bar 5}\right\}_S \otimes \ell \rangle, \label{wavefunctionspin}
\end{equation}
where $S_{12}=1$ for symmetric flavor wave function and  $S_{12}=0$ for antisymmetic wave function for the first diquark. This  pattern follows for the second diquark, where $S_{34}=1$ or $0$ for  symmetric or antisymmetric flavor wave functions, respectively. $\ell$ represents the orbital angular momentum. In this work, we assume \( \ell = 0 \), corresponding to S-wave pentaquark states, where the total orbital angular momentum includes the relative motion between the subclusters (diquark-diquark-antiquark).

\subsection{Magnetic moment}
Magnetic moments of the hadrons give valuable information about the internal structure, charge distribution and geometric shape of the hadron. As mentioned before, the Fermi statistics requires that the overall wave function of a pentaquark should be antisymmetric. In the ground state, color wave function $\epsilon_{color}$ is antisymmetric and space wave function $\eta_{spatial}$ is symmetric which yields $\epsilon_{color} \otimes \eta_{spatial}$ be antisymmetric. Therefore, one needs to consider spin-flavor wave function $\chi_{spin} \otimes \phi_{flavor}$ for the calculation of magnetic moment.

The quark configuration for the diquark-diquark-antiquark scheme is $(Qq_1)(q_2q_3)\bar{Q}$. Specifically, the diquark-diquark-antiquark configuration is $(c q_1) (q_2 q_3) (\bar{c})$ and the wave function for this state becomes 
\begin{equation}
\vert \text{Pentaquark} \rangle=\left|\left\{ \left[(cq_1)_{s_H}\otimes(q_2q_3)_{s_L}\right]_{s_{HL}}\otimes
{\bar{c}}\right\}_S\otimes \ell\right\rangle.
\end{equation}
Here $s_H$ is the spin of the $(cq_1)$ diquark, $s_L$ is the spin of the $(q_2q_3)$ diquark which couples $s_H \otimes s_L$ to the spin $s_{HL}$ of the four-quark structure. This four-quark structure then couples to the spin of anticharm $\bar{c}$ to form the total spin of the pentaquark state. $\ell$ denotes the orbital excitation. At the quark level, magnetic moment operators can be written as
\begin{equation}
\hat{\mu}_{\text{spin}}=\sum_i \frac{q_i}{2m_i}\hat{\sigma}_i,
\end{equation}
where $q_i$ is the quark charge, $m_i$ is the quark mass, and $\sigma_i$ is the Pauli's spin matrix. Each constitute may contribute to the magnetic moment of the related state. For the up, down, strange and charm quarks, we use the constituent quark masses as input \cite{Wang:2018gpl}:

\begin{equation}
m_u=m_d=0.336\ \mbox{GeV},~ m_s=0.540\ \mbox{GeV}, ~m_c=1.660\ \mbox{GeV}.
\end{equation}

\section{Numerical results and Discussion}\label{numerical}

We present $S$-wave magnetic moment results of $8_{1f}$ and $8_{2f}$ representations.  The reason of choosing these configurations is that $P_c(4380)$ and $P_c(4450)$ states with isospin $(I,I_3)=(\frac{1}{2},\frac{1}{2})$ are in the $8_{1f}$ or $8_{2f}$ representation in the diquark-diquark-antiquark scheme \cite{Wang:2016dzu}. In addition to this, $P_{cs}(4459)$ is supposed to be in $8_{2f}$ representation \cite{Gao:2021hmv}. 

Before presenting magnetic moments of the $P_c(4457)$ and its related states, we want to denote the labelling convention of the states in this work. We use a labelling convention ($P_{cri}$, where $r$ denotes for the related $P_c(4457)$ states and $i$ counts the sates) in Table \ref{labelpentaquarks8f} in order to distinguish the states\footnote{For the related $P_c(4457)$ states, we adopt the convention in Ref. \cite{Ozdem:2024usw}}.  The results are listed in Tables  \ref{result1}, \ref{result2}, \ref{result3}, \ref{result4}, \ref{result5}, \ref{result6}, for $P_c(4457)$, $P_{cr1}$, $P_{cr2}$, $P_{cr3}$, $P_{cr4}$ and $P_{cr5}$ states, respectively.

\begin{widetext}

\begin{table*}[htb]
\caption{Flavor wave functions of $P_c$ and related states in $8_f$ representation}\label{labelpentaquarks8f}
\begin{tabular}{l|c|c}
\toprule
Type of states & Wave function-$8_{1f}$& Wave function-$8_{2f}$  \\
\toprule
$P_c(4457)$&     $-\sqrt{\frac{1}{3}}\{ud\}(cu){\bar c}+\sqrt{\frac{2}{3}}\{uu\}{({c}d)\bar c}$       &   $[ud]({c}u)\bar c$          \\
$P_{cr1}$ & $\sqrt{\frac{1}{3}}\{ud\}(cd){\bar c}-\sqrt{\frac{2}{3}}\{dd\}({c}u){\bar c}$ &   $[ud](cd){\bar c}$   \\
$P_{cr2}$  & $\sqrt{\frac{1}{3}}\{us\}({c}u){\bar
c}-\sqrt{\frac{2}{3}}\{uu\}(cs){\bar c}$  &  $[us](cu){\bar c}$    \\
$P_{cr3}$  &  $ \sqrt{\frac{1}{3}}\{ds\}({c}d){\bar
c}-\sqrt{\frac{2}{3}}\{dd\}(cs){\bar c}$
 &  $[ds](cd){\bar c}$     \\
$P_{cr4}$ & $\sqrt{\frac{1}{3}}\{us\}({c}s){\bar
c}-\sqrt{\frac{2}{3}}\{ss\}({c}u){\bar c}$   &  $[us](cs){\bar c}$  \\
$P_{cr5}$  &$\sqrt{\frac{1}{3}}\{ds\}({c}s){\bar
c}-\sqrt{\frac{2}{3}}\{ss\}({c}d){\bar c}$   &  $[ds](cs){\bar c}$ \\

\toprule
\end{tabular}
\end{table*}

\end{widetext}  

\renewcommand{\arraystretch}{1.6}
\begin{table*}[h!]
\caption{The magnetic moment results of the $P_c(4457)$ pentaquark. $J_{H}^{P_H} $ corresponds to the angular momentum and parity of $(cq_1)$, $J_{L}^{P_L}$ is for $(q_2q_3)$, $J_{\bar{c}}^{P_{\bar{c}}}$ is for $\bar{c}$, and $J_{\ell}^{P_{\ell}}$ is for orbital. The results are presented in unit of nuclear magneton $\mu_N$. }\label{result1}
\begin{center}
   \begin{tabular}{c|c|c|c} \toprule[1pt]\toprule[1pt]
 \multicolumn{4}{c} {$8_{2f}$ representation}\\\toprule[1pt]
$J^P$ &$^{2s+1}L_J$ &$J_{H}^{P_H} \otimes J_{L}^{P_L}\otimes J_{\bar{c}}^{P_{\bar{c}}} \otimes J_{\ell}^{P_{\ell}}$ &Results \\\midrule[1pt]

{$\frac{1}{2}^{-}$}  &{{${^{2}S_{\frac{1}{2}}}$}}   &${0}^{+}\otimes0^{+}\otimes{\frac{1}{2}}^{-} \otimes{0^{+}}$ &-0.377\\

& &${1}^{+}\otimes0^{+}\otimes{\frac{1}{2}}^{-} \otimes{0^{+}}$ &-0.244\\

\hline
{$\frac{3}{2}^{-}$}  &{{${^{4}S_{\frac{3}{2}}}$}}   & ${1}^{+}\otimes0^{+}\otimes{\frac{1}{2}}^{-} \otimes{0^{+}}$  &-0.930\\
\hline

\bottomrule[1pt]

   \multicolumn{4}{c} {$8_{1f}$ representation}\\\toprule[1pt]
\,\,\,\,$J^P$\,\,\,\,    &\,$^{2s+1}L_J$\,  & $J_{H}^{P_H} \otimes J_{L}^{P_L}\otimes J_{\bar{c}}^{P_{\bar{c}}} \otimes J_{\ell}^{P_{\ell}}$  & Results
\\\midrule[1pt]

{$\frac{1}{2}^{-}$}  &{{${^{2}S_{\frac{1}{2}}}$}}   &$0^{+}\otimes1^{+} \otimes {\frac{1}{2}}^{-}\otimes0^{+}$ &2.607\\

&&$(1^{+}\otimes1^{+})_{0} \otimes {\frac{1}{2}}^{-}\otimes0^{+}$ &-0.377\\
&&$(1^{+}\otimes1^{+})_{1} \otimes {\frac{1}{2}}^{-}\otimes0^{+}$  &1.182\\

\hline

{$\frac{3}{2}^{-}$}  &{{${^{4}S_{\frac{3}{2}}}$}}    &$(0^{+}\otimes1^{+})\otimes{\frac{1}{2}}^{-}\otimes0^{+}$  &3.345\\

&&$(1^{+}\otimes1^{+})_{1} \otimes {\frac{1}{2}}^{-}\otimes0^{+}$ &1.207\\

& &$(1^{+}\otimes1^{+})_{2} \otimes {\frac{1}{2}}^{-}\otimes0^{+}$  &3.077\\
\hline

\multirow{1}{*}{$\frac{5}{2}^{-}$}  &{\multirow{1}*{${^{6}S_{\frac{5}{2}}}$}} &$1^{+}\otimes1^{+}\otimes{\frac{1}{2}}^{-}\otimes0^{+}$  &2.792\\

\hline

     \bottomrule[1pt]\bottomrule[1pt]
      \end{tabular}
  \end{center}
\end{table*}

\renewcommand{\arraystretch}{1.6}
\begin{table*}[h!]
\caption{Same as in Table \ref{result1} but for the the $P_{cr1}$ pentaquark.}\label{result2}
\begin{center}
   \begin{tabular}{c|c|c|c} \toprule[1pt]\toprule[1pt]
 \multicolumn{4}{c} {$8_{2f}$ representation}\\\toprule[1pt]
$J^P$ &$^{2s+1}L_J$ &$J_{H}^{P_H} \otimes J_{L}^{P_L}\otimes J_{\bar{c}}^{P_{\bar{c}}} \otimes J_{\ell}^{P_{\ell}}$ &Results \\\midrule[1pt]

{$\frac{1}{2}^{-}$}  &{{${^{2}S_{\frac{1}{2}}}$}}   &${0}^{+}\otimes0^{+}\otimes{\frac{1}{2}}^{-} \otimes{0^{+}}$ &-0.377\\

& &${1}^{+}\otimes0^{+}\otimes{\frac{1}{2}}^{-} \otimes{0^{+}}$ &1.617\\

\hline
{$\frac{3}{2}^{-}$}  &{{${^{4}S_{\frac{3}{2}}}$}}   & ${1}^{+}\otimes0^{+}\otimes{\frac{1}{2}}^{-} \otimes{0^{+}}$  &1.861\\
\hline

\bottomrule[1pt]

   \multicolumn{4}{c} {$8_{1f}$ representation}\\\toprule[1pt]
\,\,\,\,$J^P$\,\,\,\,    &\,$^{2s+1}L_J$\,  & $J_{H}^{P_H} \otimes J_{L}^{P_L}\otimes J_{\bar{c}}^{P_{\bar{c}}} \otimes J_{\ell}^{P_{\ell}}$  & Results
\\\midrule[1pt]

{$\frac{1}{2}^{-}$}  &{{${^{2}S_{\frac{1}{2}}}$}}   &$0^{+}\otimes1^{+} \otimes {\frac{1}{2}}^{-}\otimes0^{+}$ &-1.115\\

&&$(1^{+}\otimes1^{+})_{0} \otimes {\frac{1}{2}}^{-}\otimes0^{+}$ & -0.377\\
&&$(1^{+}\otimes1^{+})_{1} \otimes {\frac{1}{2}}^{-}\otimes0^{+}$  & 0.251 \\

\hline

{$\frac{3}{2}^{-}$}  &{{${^{4}S_{\frac{3}{2}}}$}}    &$(0^{+}\otimes1^{+})\otimes{\frac{1}{2}}^{-}\otimes0^{+}$  & -2.238\\

&&$(1^{+}\otimes1^{+})_{1} \otimes {\frac{1}{2}}^{-}\otimes0^{+}$ &-0.188\\

& &$(1^{+}\otimes1^{+})_{2} \otimes {\frac{1}{2}}^{-}\otimes0^{+}$  & 0.565\\
\hline

\multirow{1}{*}{$\frac{5}{2}^{-}$}  &{\multirow{1}*{${^{6}S_{\frac{5}{2}}}$}} &$1^{+}\otimes1^{+}\otimes{\frac{1}{2}}^{-}\otimes0^{+}$  &0\\

\hline
     \bottomrule[1pt]\bottomrule[1pt]
      \end{tabular}
  \end{center}
\end{table*}

\renewcommand{\arraystretch}{1.6}
\begin{table*}[h!]
\caption{Same as in Table \ref{result1} but for the the $P_{cr2}$ pentaquark.}\label{result3}
\begin{center}
   \begin{tabular}{c|c|c|c} \toprule[1pt]\toprule[1pt]
 \multicolumn{4}{c} {$8_{2f}$ representation}\\\toprule[1pt]
$J^P$ &$^{2s+1}L_J$ &$J_{H}^{P_H} \otimes J_{L}^{P_L}\otimes J_{\bar{c}}^{P_{\bar{c}}} \otimes J_{\ell}^{P_{\ell}}$ &Results \\\midrule[1pt]

{$\frac{1}{2}^{-}$}  &{{${^{2}S_{\frac{1}{2}}}$}}   &${0}^{+}\otimes0^{+}\otimes{\frac{1}{2}}^{-} \otimes{0^{+}}$ &-0.377\\

& &${1}^{+}\otimes0^{+}\otimes{\frac{1}{2}}^{-} \otimes{0^{+}}$ &-0.009 \\

\hline
{$\frac{3}{2}^{-}$}  &{{${^{4}S_{\frac{3}{2}}}$}}   & ${1}^{+}\otimes0^{+}\otimes{\frac{1}{2}}^{-} \otimes{0^{+}}$  &-0.579 \\
\hline

\bottomrule[1pt]

   \multicolumn{4}{c} {$8_{1f}$ representation}\\\toprule[1pt]
\,\,\,\,$J^P$\,\,\,\,    &\,$^{2s+1}L_J$\,  & $J_{H}^{P_H} \otimes J_{L}^{P_L}\otimes J_{\bar{c}}^{P_{\bar{c}}} \otimes J_{\ell}^{P_{\ell}}$  & Results
\\\midrule[1pt]

{$\frac{1}{2}^{-}$}  &{{${^{2}S_{\frac{1}{2}}}$}}   &$0^{+}\otimes1^{+} \otimes {\frac{1}{2}}^{-}\otimes0^{+}$ & 2.607 \\

&&$(1^{+}\otimes1^{+})_{0} \otimes {\frac{1}{2}}^{-}\otimes0^{+}$ & -0.377\\
&&$(1^{+}\otimes1^{+})_{1} \otimes {\frac{1}{2}}^{-}\otimes0^{+}$  & 1.299 \\

\hline

{$\frac{3}{2}^{-}$}  &{{${^{4}S_{\frac{3}{2}}}$}}    &$(0^{+}\otimes1^{+})\otimes{\frac{1}{2}}^{-}\otimes0^{+}$  & 3.345 \\

&&$(1^{+}\otimes1^{+})_{1} \otimes {\frac{1}{2}}^{-}\otimes0^{+}$ &1.383 \\

& &$(1^{+}\otimes1^{+})_{2} \otimes {\frac{1}{2}}^{-}\otimes0^{+}$  &3.394 \\
\hline

\multirow{1}{*}{$\frac{5}{2}^{-}$}  &{\multirow{1}*{${^{6}S_{\frac{5}{2}}}$}} &$1^{+}\otimes1^{+}\otimes{\frac{1}{2}}^{-}\otimes0^{+}$  & 3.143\\

\hline
     \bottomrule[1pt]\bottomrule[1pt]
      \end{tabular}
  \end{center}
\end{table*}

\renewcommand{\arraystretch}{1.6}
\begin{table*}[h!]
\caption{Same as in Table \ref{result1} but for the the $P_{cr3}$ pentaquark.}\label{result4}
\begin{center}
   \begin{tabular}{c|c|c|c} \toprule[1pt]\toprule[1pt]
 \multicolumn{4}{c} {$8_{2f}$ representation}\\\toprule[1pt]
$J^P$ &$^{2s+1}L_J$ &$J_{H}^{P_H} \otimes J_{L}^{P_L}\otimes J_{\bar{c}}^{P_{\bar{c}}} \otimes J_{\ell}^{P_{\ell}}$ &Results \\\midrule[1pt]

{$\frac{1}{2}^{-}$}  &{{${^{2}S_{\frac{1}{2}}}$}}   &${0}^{+}\otimes0^{+}\otimes{\frac{1}{2}}^{-} \otimes{0^{+}}$ &-0.377\\

& &${1}^{+}\otimes0^{+}\otimes{\frac{1}{2}}^{-} \otimes{0^{+}}$ &-0.009 \\

\hline
{$\frac{3}{2}^{-}$}  &{{${^{4}S_{\frac{3}{2}}}$}}   & ${1}^{+}\otimes0^{+}\otimes{\frac{1}{2}}^{-} \otimes{0^{+}}$  & -0.579 \\
\hline

\bottomrule[1pt]

   \multicolumn{4}{c} {$8_{1f}$ representation}\\\toprule[1pt]
\,\,\,\,$J^P$\,\,\,\,    &\,$^{2s+1}L_J$\,  & $J_{H}^{P_H} \otimes J_{L}^{P_L}\otimes J_{\bar{c}}^{P_{\bar{c}}} \otimes J_{\ell}^{P_{\ell}}$  & Results
\\\midrule[1pt]

{$\frac{1}{2}^{-}$}  &{{${^{2}S_{\frac{1}{2}}}$}}   &$0^{+}\otimes1^{+} \otimes {\frac{1}{2}}^{-}\otimes0^{+}$ & -1.115 \\

&&$(1^{+}\otimes1^{+})_{0} \otimes {\frac{1}{2}}^{-}\otimes0^{+}$ & -0.377 \\
&&$(1^{+}\otimes1^{+})_{1} \otimes {\frac{1}{2}}^{-}\otimes0^{+}$  & -0.562 \\

\hline

{$\frac{3}{2}^{-}$}  &{{${^{4}S_{\frac{3}{2}}}$}}    &$(0^{+}\otimes1^{+})\otimes{\frac{1}{2}}^{-}\otimes0^{+}$  & -2.238 \\

&&$(1^{+}\otimes1^{+})_{1} \otimes {\frac{1}{2}}^{-}\otimes0^{+}$ &-1.408 \\

& &$(1^{+}\otimes1^{+})_{2} \otimes {\frac{1}{2}}^{-}\otimes0^{+}$ &-1.631 \\
\hline

\multirow{1}{*}{$\frac{5}{2}^{-}$}  &{\multirow{1}*{${^{6}S_{\frac{5}{2}}}$}} &$1^{+}\otimes1^{+}\otimes{\frac{1}{2}}^{-}\otimes0^{+}$  & -2.440 \\

\hline
     \bottomrule[1pt]\bottomrule[1pt]
      \end{tabular}
  \end{center}
\end{table*}

\renewcommand{\arraystretch}{1.6}
\begin{table*}[h!]
\caption{Same as in Table \ref{result1} but for the the $P_{cr4}$ pentaquark.}\label{result5}
\begin{center}
   \begin{tabular}{c|c|c|c} \toprule[1pt]\toprule[1pt]
 \multicolumn{4}{c} {$8_{2f}$ representation}\\\toprule[1pt]
$J^P$ &$^{2s+1}L_J$ &$J_{H}^{P_H} \otimes J_{L}^{P_L}\otimes J_{\bar{c}}^{P_{\bar{c}}} \otimes J_{\ell}^{P_{\ell}}$ &Results \\\midrule[1pt]

{$\frac{1}{2}^{-}$}  &{{${^{2}S_{\frac{1}{2}}}$}}   &${0}^{+}\otimes0^{+}\otimes{\frac{1}{2}}^{-} \otimes{0^{+}}$ & -0.377 \\

& &${1}^{+}\otimes0^{+}\otimes{\frac{1}{2}}^{-} \otimes{0^{+}}$ & 1.617 \\

\hline
{$\frac{3}{2}^{-}$}  &{{${^{4}S_{\frac{3}{2}}}$}}   & ${1}^{+}\otimes0^{+}\otimes{\frac{1}{2}}^{-} \otimes{0^{+}}$  & 1.861 \\
\hline

\bottomrule[1pt]

   \multicolumn{4}{c} {$8_{1f}$ representation}\\\toprule[1pt]
\,\,\,\,$J^P$\,\,\,\,    &\,$^{2s+1}L_J$\,  & $J_{H}^{P_H} \otimes J_{L}^{P_L}\otimes J_{\bar{c}}^{P_{\bar{c}}} \otimes J_{\ell}^{P_{\ell}}$  & Results
\\\midrule[1pt]

{$\frac{1}{2}^{-}$}  &{{${^{2}S_{\frac{1}{2}}}$}}   &$0^{+}\otimes1^{+} \otimes {\frac{1}{2}}^{-}\otimes0^{+}$ & -0.646 \\

&&$(1^{+}\otimes1^{+})_{0} \otimes {\frac{1}{2}}^{-}\otimes0^{+}$ & -0.377 \\
&&$(1^{+}\otimes1^{+})_{1} \otimes {\frac{1}{2}}^{-}\otimes0^{+}$  & 0.485 \\

\hline

{$\frac{3}{2}^{-}$}  &{{${^{4}S_{\frac{3}{2}}}$}}    &$(0^{+}\otimes1^{+})\otimes{\frac{1}{2}}^{-}\otimes0^{+}$  & -1.535 \\

&&$(1^{+}\otimes1^{+})_{1} \otimes {\frac{1}{2}}^{-}\otimes0^{+}$ &0.163 \\

& &$(1^{+}\otimes1^{+})_{2} \otimes {\frac{1}{2}}^{-}\otimes0^{+}$ & 1.198 \\
\hline

\multirow{1}{*}{$\frac{5}{2}^{-}$}  &{\multirow{1}*{${^{6}S_{\frac{5}{2}}}$}} &$1^{+}\otimes1^{+}\otimes{\frac{1}{2}}^{-}\otimes0^{+}$  & 0.703 \\

\hline
     \bottomrule[1pt]\bottomrule[1pt]
      \end{tabular}
  \end{center}
\end{table*}

\renewcommand{\arraystretch}{1.6}
\begin{table*}[h!]
\caption{Same as in Table \ref{result1} but for the the $P_{cr5}$ pentaquark.}\label{result6}
\begin{center}
   \begin{tabular}{c|c|c|c} \toprule[1pt]\toprule[1pt]
 \multicolumn{4}{c} {$8_{2f}$ representation}\\\toprule[1pt]
$J^P$ &$^{2s+1}L_J$ &$J_{H}^{P_H} \otimes J_{L}^{P_L}\otimes J_{\bar{c}}^{P_{\bar{c}}} \otimes J_{\ell}^{P_{\ell}}$ &Results \\\midrule[1pt]

{$\frac{1}{2}^{-}$}  &{{${^{2}S_{\frac{1}{2}}}$}}   &${0}^{+}\otimes0^{+}\otimes{\frac{1}{2}}^{-} \otimes{0^{+}}$ & -0.377 \\

& &${1}^{+}\otimes0^{+}\otimes{\frac{1}{2}}^{-} \otimes{0^{+}}$ &  -0.244  \\

\hline
{$\frac{3}{2}^{-}$}  &{{${^{4}S_{\frac{3}{2}}}$}}   & ${1}^{+}\otimes0^{+}\otimes{\frac{1}{2}}^{-} \otimes{0^{+}}$  & -0.930  \\
\hline

\bottomrule[1pt]

   \multicolumn{4}{c} {$8_{1f}$ representation}\\\toprule[1pt]
\,\,\,\,$J^P$\,\,\,\,    &\,$^{2s+1}L_J$\,  & $J_{H}^{P_H} \otimes J_{L}^{P_L}\otimes J_{\bar{c}}^{P_{\bar{c}}} \otimes J_{\ell}^{P_{\ell}}$  & Results
\\\midrule[1pt]

{$\frac{1}{2}^{-}$}  &{{${^{2}S_{\frac{1}{2}}}$}}   &$0^{+}\otimes1^{+} \otimes {\frac{1}{2}}^{-}\otimes0^{+}$ & -0.646 \\

&&$(1^{+}\otimes1^{+})_{0} \otimes {\frac{1}{2}}^{-}\otimes0^{+}$ & -0.377\\
&&$(1^{+}\otimes1^{+})_{1} \otimes {\frac{1}{2}}^{-}\otimes0^{+}$  &  -0.445 \\

\hline

{$\frac{3}{2}^{-}$}  &{{${^{4}S_{\frac{3}{2}}}$}}    &$(0^{+}\otimes1^{+})\otimes{\frac{1}{2}}^{-}\otimes0^{+}$  & -1.535 \\

&&$(1^{+}\otimes1^{+})_{1} \otimes {\frac{1}{2}}^{-}\otimes0^{+}$ &-1.233   \\

& &$(1^{+}\otimes1^{+})_{2} \otimes {\frac{1}{2}}^{-}\otimes0^{+}$ &-1.315 \\
\hline

\multirow{1}{*}{$\frac{5}{2}^{-}$}  &{\multirow{1}*{${^{6}S_{\frac{5}{2}}}$}} &$1^{+}\otimes1^{+}\otimes{\frac{1}{2}}^{-}\otimes0^{+}$  & -2.088 \\

\hline
     \bottomrule[1pt]\bottomrule[1pt]
      \end{tabular}
  \end{center}
\end{table*}

Based on the predictions presented, we highlight our results as follows:

\begin{itemize}

\item The \( J^P \) assignments in Tables \ref{result1}–\ref{result6} follow the conventions: the anti-charm quark has \( J^P = 1/2^- \) due to its fermionic nature and negative intrinsic parity; the orbital part is assumed to be in an S-wave (\( J^P = 0^+ \)) throughout this work.

\item As can be seen in the tables, all the states in the ${0}^{+}\otimes0^{+}\otimes{\frac{1}{2}}^{-} \otimes{0^{+}}$  configuration of $J^P=\frac{1}{2}^{-}$ quantum number in $8_{2f}$ flavor representation and $(1^{+}\otimes1^{+})_{0} \otimes {\frac{1}{2}}^{-}\otimes0^{+}$ configuration of $J^P=\frac{1}{2}^{-}$ quantum number in $8_{1f}$ flavor representation have same magnetic moments, $\mu=-0.377 \ \mu_N$. This is the magnetic moment of the anticharm quark. Therefore, only anticharm quark contributes to the pentaquark magnetic moment in this case. 

\item In $P_c(4457)$ state, the magnetic moments of the $8_{1f}$ and $8_{2f}$ flavor representations are different, except the two cases mentioned above. In $8_{2f}$ representation, all the magnetic moments are negative whereas except $(1^{+}\otimes1^{+})_{0} \otimes {\frac{1}{2}}^{-}\otimes0^{+}$ with  $J^P=\frac{1}{2}^{-}$ in $8_{1f}$ representation, all the magnetic moments are positive. In Ref. \cite{Ozdem:2021ugy}, magnetic moment of $P_c(4457)$ with $\frac{1}{2}^{-}$ quantum number was studied by using light-cone QCD sum rule formalism in diquark-diquark-antiquark and molecular pictures where the results for magnetic moments are $\mu_{P_c({4457})}=0.88^{+0.32}_{-0.29} \ \mu_N$. Our result of $J^P=\frac{1}{2}^{-}$ $(1^{+}\otimes1^{+})_{1} \otimes {\frac{1}{2}}^{-}\otimes0^{+}$  in $8_{1f}$ representation is $\mu_N=1.182  \ \mu_N$ and compatible with the prediction of reference work. Using quark model in a framework with and without coupled channel and D-wave effects \cite{Li:2021ryu}, magnetic moment of $P_c(4457)$ with $J^P=\frac{3}{2}^{-}$ quantum number was obtained as $\mu_{P_c({4457})}=(1.145-1.365)\ \mu_N$. Our result of $J^P=\frac{3}{2}^{-}$ $(1^{+}\otimes1^{+})_{1} \otimes {\frac{1}{2}}^{-}\otimes0^{+}$  in $8_{1f}$ representation is $\mu=1.207 \ \mu_N$ and agree well with the given reference. In Ref. \cite{Ozdem:2024usw}, magnetic moment of $P_c(4457)$ with $J^P=\frac{3}{2}^{-}$ in diquark-diquark-antiquark scheme is obtained as $\mu_{P_c({4457})}=-1.96^{+0.50}_{-0.37} \ \mu_N$ by using light-cone QCD sum rule formalism. The results of $8_{2f}$ representation are negative but are not compatible with this result in terms of magnitude. The result of  ${1}^{+}\otimes0^{+}\otimes{\frac{1}{2}}^{-} \otimes{0^{+}}$ of $J^P=\frac{3}{2}^{-}$  in $8_{2f}$ representation is  $\mu=-0.930  \ \mu_N$ and close to the value of given reference within the up and down limits.

\item In the $P_{cr1}$ type pentaquark, magnetic moment of the $1^{+}\otimes1^{+}\otimes{\frac{1}{2}}^{-}\otimes0^{+}$ configuration with $J^P=\frac{5}{2}^{-}$ quantum number turns out to be zero. Both $J^P=\frac{1}{2}^{-}$ and $J^P=\frac{3}{2}^{-}$ quantum numbers in $8_{1f}$ and $8_{2f}$ representations are quite different and a possible observation of magnetic moment may help to identify quantum number of this state. Ref. \cite{Ozdem:2024usw} obtained magnetic moment of this state with $J^P=\frac{3}{2}^{-}$ quantum number as $\mu_N=-2.04^{+0.46}_{-0.41} \ \mu_N$ using light-cone QCD sum rule formalism. Our result of  $J^P=\frac{3}{2}^{-}$ $(0^{+}\otimes1^{+})\otimes{\frac{1}{2}}^{-}\otimes0^{+}$ in $8_{1f}$ representation is  $\mu=-2.238 \ \mu_N$ and agree with this result. Compared to $P_c(4457)$ state, it can be seen that quark rearrangements in diquarks drastically change the magnetic moment results. 

\item In the $P_{cr2}$ type pentaquark, all the results in $8_{2f}$ representation are negative. In the $8_{1f}$ representation, except $(1^{+}\otimes1^{+})_{0} \otimes {\frac{1}{2}}^{-}\otimes0^{+}$ with  $J^P=\frac{1}{2}^{-}$, all the results are positive. None of the our results agree with the prediction in Ref. \cite{Ozdem:2024usw} which is $\mu=-2.08^{+0.53}_{-0.39} \ \mu_N$ for $J^P=\frac{3}{2}^{-}$ quantum number. 

\item In the $P_{cr3}$ type pentaquark, all the results of magnetic moments are negative in both $8_{1f}$ and $8_{2f}$ representations.  Ref. \cite{Ozdem:2024usw} obtained magnetic moment as $\mu_N=-2.13^{+0.53}_{-0.40} \ \mu_N$ for $J^P=\frac{3}{2}^{-}$ for quantum number. Our results for $J^P=\frac{3}{2}^{-}$  $(0^{+}\otimes1^{+})\otimes{\frac{1}{2}}^{-}\otimes0^{+}$ configuration is $\mu=-2.238 \ \mu_N$  agree well the result of reference value. 

\item In the $P_{cr4}$ type pentaquark, the magnetic  moment results scatter between negative and positive values. Ref. \cite{Ozdem:2024usw} obtained magnetic moment as $\mu=-2.29^{+0.53}_{-0.39} \ \mu_N$ for $J^P=\frac{3}{2}^{-}$ quantum number.  Our result for $J^P=\frac{3}{2}^{-}$  $(0^{+}\otimes1^{+})\otimes{\frac{1}{2}}^{-}\otimes0^{+}$ configuration is $\mu=-1.535 \ \mu_N$ which is compatible. 

\item In the $P_{cr5}$ type pentaquark, all the results of magnetic moments are negative in both $8_{1f}$ and $8_{2f}$ representations. Our result of $0^{+}\otimes1^{+}\otimes{\frac{1}{2}}^{-}\otimes0^{+}$ configuration with $J^P=\frac{3}{2}^{-}$ is compatible the result of Ref. \cite{Ozdem:2024usw}, which is $\mu=-2.33^{+0.53}_{-0.41} \ \mu_N$.

\item In $8_{2f}$ representations,  ${0}^{+}\otimes0^{+}\otimes{\frac{1}{2}}^{-} \otimes{0^{+}}$ configuration with $J^P=\frac{1}{2}^{-}$ and  ${1}^{+}\otimes0^{+}\otimes{\frac{1}{2}}^{-} \otimes{0^{+}}$ configuration with $J^P=\frac{3}{2}^{-}$ of $P_c(4457)$ and $P_{cr5}$, $P_{cr1}$ and $P_{cr4}$, $P_{cr2}$ and $P_{cr3}$ states have the same magnetic moments. The reason for this could be that the contribution of light quarks to the total magnetic moment is dominant. 

\item In $8_{1f}$ representations, $0^{+}\otimes1^{+} \otimes {\frac{1}{2}}^{-}\otimes0^{+}$ configuration with $J^P=\frac{1}{2}^{-}$ and $(0^{+}\otimes1^{+}) \otimes {\frac{1}{2}}^{-} \otimes 0^{+}$ configuration with $J^P=\frac{3}{2}^{-}$ of $P_c(4457)$ and $P_{cr2}$, $P_{cr1}$ and $P_{cr3}$, $P_{cr4}$ and $P_{cr5}$ states have the same magnetic moments. The reason for this could be that the contribution of light quarks to the total magnetic moment is dominant. 

\item The results of present study together with the available results in the literature magnetic moments of the hidden-charm pentaquark states may give information about internal structure and their spin-parity quantum numbers. These predictions can help distinguish between various theoretical models based on experimental measurements.

\end{itemize}

\section{Summary}\label{summary}

In this present study, we systematically study magnetic moments of the $P_{c}(4457)$ and its related hidden-charm pentaquark states with and without strangeness employing a comprehensive analysis that encompasses diquark-diquark-antiquark scheme with $J^P = \frac{1}{2}^-$, $J^P = \frac{3}{2}^-$ and $J^P = \frac{5}{2}^-$ quantum numbers. We also compare our results with the available results reported in the literature. 

The magnetic moments of hidden-charm pentaquark states vary with phenomenological models. These variations can help distinguish between different structural models such as molecular, diquark-diquark-antiquark, and diquark-triquark models. Different  phenomenological models predict distinct magnetic moments for hidden-charm pentaquark states. These predictions can help distinguish between various theoretical models based on experimental measurements. It is clear that the experimental measurement of the magnetic moment of the pentaquarks can help distinguish their inner structure. We hope that our results will help the endeavour for the probing the internal sructure of $P_c(4457)$ and its related pentaquarks.

\bibliographystyle{elsarticle-num}
\bibliography{Pc4457related-revised.bib}

\end{document}